\documentclass[runningheads]{llncs}
\usepackage{graphicx}
\usepackage{longtable}
\usepackage{booktabs}
\usepackage{lscape}
\usepackage[table,xcdraw]{xcolor}
\begin{document}
\title{Design of an Extended Reality Collaboration Architecture for Mixed Immersive and Multi-Surface Interaction}

% Design a XR multi-system architecture to analyze maritime data
% A XR multi-system collaborative framework for decision-making process in naval organizations

%Design a XR multi-system for data visualization analysis and decision-making process in naval organizations

%
\titlerunning{Design of an XR Collab. Arch. for Mixed Immersive and MS Interaction}
\author{Thiago Porcino\orcidID{0000-0002-0281-4580} \and
Seyed Adel Ghaeinian\orcidID{0000-0002-2043-1604} \and
Juliano Franz\orcidID{0000-0003-1196-0210} \and
Joseph Malloch\orcidID{0000-0001-9684-2269} \and
Derek Reilly\orcidID{0000-0003-4960-8757}
}

% \author{Thiago Porcino\inst{1} \and
% Adel Ghaeinian\inst{1} \and
% Juliano Franz\inst{1} \and
% Joseph Malloch\inst{1} \and
% Derek Reilly\inst{1}}
% }
% %
\authorrunning{T. Porcino et al.}
\institute{Graphics and Experiential Media Lab, Dalhousie University, Halifax NS, Canada\\ 
\url{https://gem.cs.dal.ca/}\\ 
\email{\{thiago,adelghaeinian\}@dal.ca}}

% \email{\{thiago,adelghaeinian,juliano.franz,jmalloch\}@dal.ca},\{reilly\}@cs.dal.ca}
%
\maketitle              % typeset the header of the contribution
\begin{abstract}
EXtended Reality (XR) is a rapidly developing paradigm for computer entertainment, and is also increasingly used for simulation, training, data analysis, and other non-entertainment purposes, often employing head-worn XR devices like the Microsoft HoloLens. In XR, integration with the physical world should also include integration with commonly used digital devices. This paper proposes an architecture to integrate head-worn displays with touchscreen devices, such as phones, tablets, or large tabletop or wall displays. The architecture emerged through the iterative development of a prototype for collaborative analysis and decision-making for the maritime domain. However, our architecture can flexibly support a range of domains and purposes.
XR designers, XR entertainment researchers, and game developers can benefit from our architecture to propose new ways of gaming, considering multiple devices as user interfaces in an immersive collaborative environment.

\end{abstract}
\section{Introduction}

We have seen rapid development and increased public interest in mixed reality technologies in recent years \cite{calvelo2020immersive} in areas such as training, analytics, and entertainment. In 2021, the augmented reality (AR), virtual reality (VR), and mixed reality (MR) market was valued at 30.7 billion U.S. dollars and it is expected to reach close to 300 billion by 2024 \cite{statista2020statistics}.

Head-worn displays (HWDs) are one way to achieve immersive mixed reality or virtual reality. These devices usually consist of electronic displays and lenses that are fixed over the head toward the eyes of the user. HWDs are used in various domains including games and entertainment \cite{studios2015elder}, military \cite{rizzo2011virtual}, education \cite{ahir2020application}, therapy \cite{carrion2019developing}, and medicine \cite{kuhnapfel2000endoscopic}. 

While some HWDs are themselves mobile devices (in that they are not tethered to a desktop PC), there is a lack of tools that assist in integrating augmented reality HWDs (such as HoloLens) and other single-user mobile devices (like phones or tablets) with shared devices such as large wall or tabletop displays. In this paper, we present a novel architecture that integrates multiple AR HWDs with a shared large display (a tabletop display in our prototype) and with handheld touchscreen devices such as  phones or tablets. AR HWDs are a natural component of multi-system collaborative environments, as they can provide personalized and/or \emph{ad hoc} extensions to interaction and visualization: for example, a visual representation can be extended beyond or above a tabletop display, gestures and head orientation can become part of a shared interface, and 3D representations of inter-device communication become possible.

While the focus of the developed prototype in this work is geospatial data analysis and visualization within a collaborative decision making context, the proposed architecture readily extends to other areas, and in particular entertain\-ment-focused XR applications such as games that integrate with shared displays, such as an interactive mixed reality board game or puzzle. 

Furthermore, our architecture allows researchers to prototype and study visualizations and interaction techniques that integrate modern HWDs and interactive touchscreens in the general case or in applied contexts such as entertainment, games, training and simulation, immersive analytics, and more.

This paper is organized as follows: Section 2 describes the related work; Section 3 details the iterative design and development of our prototype, leading to the proposed architecture. We then outline future work and current limitations in Section 4, and conclude in Section 5.

% \section{Extended Reality (XR) Background}

% By the definition extended reality (XR) is a umbrella term for virtual reality (VR), augmented reality (AR), and mixed reality (MR). 

% Virtual Reality (VR) 

\section{Related Work}

Multi-system collaborative environments require flexibility of computation, data, interaction, and visualization \cite{herz2012addressing,powers2013key}. The general idea is that devices should be dynamically reconfigurable according to a desired or emergent objective. Any multi-system collaborative environment that incorporates AR HWDs should allow for these devices to be flexibly added to, removed from, and configured for the system. 

% Gjerlufen et al.\cite{gjerlufsen2011shared} developed the programming framework ``Substance'' based on a data-oriented model where state and behavior are loosely coupled. Additionally, they produced middleware that implements a distributed application model and provides sharing abstractions, called Shared Substance. 

% Chapuis et al. \cite{chapuis2014smarties} produced the Smarties system. Smarties specializes in interactive support for multi-device collaborative applications, focusing on smart devices and wall displays. It includes an interface developed for mobile devices for inputting commands. The communication protocol handles synchronization, locking, and input conflicts and implementations are provided for C++, Java, and C\# for Unity3D. Device interfaces are decoupled, such that for example personalization of wall display applications do not require modifications to the mobile device controller application.

Salimian et al. \cite{salimian2018imrce} proposed IMRCE, a Unity toolkit for immersive mixed reality collaboration. In this work, users interact with shared 3D virtual objects using touchscreens, in VR, or in mixed reality configurations. IMRCE connects mixed groups of collocated and remote collaborators, and was designed to support rapid prototyping of mixed reality collaborative environments that use hand and position tracking as data. IMRCE was evaluated with groups of developers who developed simple prototypes using the toolkit, and with end users who performed collaborative tasks using it\cite{salimian2019MPRemix}. While IMRCE was compared against a base Unity development environment, other competitive toolkits such as TwinSpace \cite{reilly2010twinspace}, SoD-Toolkit \cite{seyed2015sod}, KinectArms \cite{genest2013kinectarms}, or Proximity \cite{marquardt2011proximity}, were not considered, in part due to IMRCE's embedding within Unity3D.

Huh et al. \cite{huh2019xr} introduce an architecture in which multiple AR/VR clients can collaborate in a shared workspace in a decentralized manner. 
Their architecture has two data categories: the data stored in the database (based data) and shared among users, and the user's data (extension data), which are a modified version of the standard data common to all users. Their architecture facilitates immersive XR collaboration between clients while the network connection is not stable enough using distributed databases and decentralized web technologies.

Ran et al. \cite{ran2020multi} focus on rendering virtual objects around a common virtual point (in the virtual world). They developed a system called SPAR (Spatially Consistent AR). According to the authors, AR apps have issues of high latency, spatial drift, and spatial inconsistency of the virtual assets distributed over time and users. 
In other words, the virtual objects aren't rendered simultaneously and with the exact position for all involved users. For this reason, SPAR attends as a new method for communication and computation of AR devices. They worked with an open-source AR platform (Android) instead of closed popular ones such as Apple and Microsoft because their code cannot be changed for tests. Authors mentioned they decreased the total latency by 55\% and spatial inconsistency by up to 60\% compared to the communication of off-the-shelf AR systems.

In summary, the related works take different approaches to integrate systems for collaboration. IMRCE \cite{salimian2018imrce}, SPAR \cite{ran2020multi}, and Huh's architecture \cite{huh2019xr} are the three works closest to our own. However, in IMRCE \cite{salimian2018imrce} content is limited to presentation in virtual reality, while in SPAR \cite{ran2020multi} immersive augmented reality content was not integrated with other devices, such as a tabletop display. Our platform supports both diverse hardware platforms and diverse forms of content and channels of content presentation. Furthermore, Huh et al. \cite{huh2019xr} designed a decentralized XR architecture quite similar to our architecture concerning AR processing and visualization. However, they do not include shared screens in their architecture. Unlike other works, in our system, collaborators can work around a shared display (SD) or use tablets, HWDs, or screen touchable devices.

\section{Materials and Methods}
The software framework presented in this paper was originally developed to support collaborative monitoring, analysis, and decision-making involving maritime data.  
The resulting system combines a shared tabletop display with tablet displays and AR HWDs used by multiple users. A cloud-based repository provides data from multiple sources to our interfaces. These data are used to generate 2D geospatial visualizations  presented on the tabletop display, which are augmented by 3D visualizations in augmented reality. Tablet displays permit individual collaborators to query and constrain the data visualization on both their personal HWD and the shared tabletop display. The tablet interface lets individuals query structured data visually, either through touch-based interaction with query primitives on the tablet, interactive selection and filtering of the geospatial visualization on the tabletop, or direct manipulation of 3D objects via the AR HWD. The querying system is built on top of SPARQL \cite{perez2009semantics}, a query language for structured databases commonly used by data-driven AI systems.  

We used the Unity 3D game engine as the primary development framework to produce the XR content. The first reason was that Unity 3D is a powerful game engine that allows the creation of rich interactive AR/MR/VR/XR experiences (such as games or simulations) for mobile devices. Second, we adopt the idea of exploring the multi-player gaming concept for asset communication between the server and clients in our AR solution. We used a library for multi-player network communication (Photon Engine) and developed the server-side (Shared Server or SS) and a client-side (AR Room). The AR Room can be described as a set of clients that can connect to the server or Shared Server (SS).

Moreover, we propose the following architecture (illustrated in Figure 1) to integrate the SD with multiple HoloLens' and tablets. The developed applications were constructed using Unity 3D as game engine for 3D models, interaction, and AR communication, which includes Photon Engine for AR multi-player behaviors between SS and clients inside the AR Room. Moreover, we used React \cite{gackenheimer2015introduction} and Leaflet \cite{crickard2014leaflet} for the web-based SD application and SPARQL for database access.  We divided our architecture in 5 entities.

% , making them compatible with Microsoft HoloLens, an HWD (head-worn display) specific for augmented reality visualizations.

\begin{figure}[ht]
\centering
\label{Fig:frame}
\centering
\includegraphics[width=0.85\textwidth]{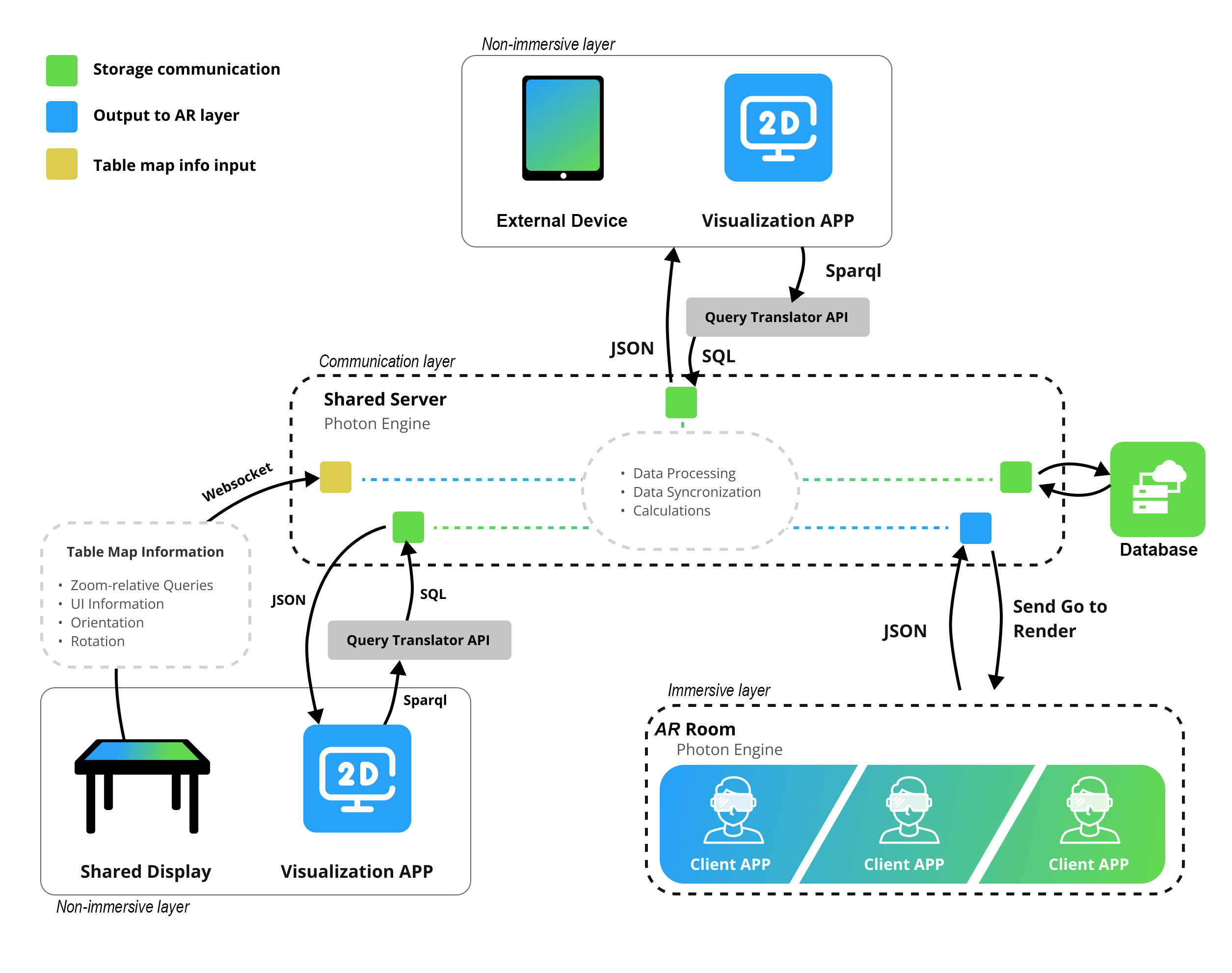}
\caption{The overall architecture. Data arrives from the right (Database), and is processed by Shared Server (SS). The user can interact with the system via a physical Shared Display (SD) (by physical touching gestures, such as clicking, pinching, pan, zooming) or augmented reality (making gestures on air in a AR context and wearing a HoloLens). AR devices (HoloLens) and new non-immersive devices (such as tablets or smartphones) can connect to the SS and see data in XR or non-immersive visualizations.} 
\end{figure} 

\begin{itemize}
    \item \textbf{Shared Display (SD)} - The SD entity represents a shared physical display with the visualization application embedded. The SD has a web-based application developed in React using the Mapbox API \cite{cadenas2014geovisualization} to show the Map and AR 3D objects. The SD sends the Map's latitude/longitude boundaries, zoom, orientation, and rotation information to the SS in real-time (each interaction in the SD is shared among other connected devices). The SD can also display a unique QR code for calibrating the coordinate systems used by the augmented reality layer (Illustrated in the "AR Room" box in Figure 1). The SD application can customize the visualization by selecting specific data using a visual query builder to consult the data source through SS. 
    \item \textbf{Shared Server (SS)} - the SS is responsible for converting lat/long data received from the database to the relative position in AR context, processing these data, transforming the selection into a structured data file (JSON), and sharing this data with other devices connected to the SS' network (i.e., Table, Database, and Augmented Reality Content). The SS also receives input queries from the SD  (Figure 1) and any other connected devices and if necessary converts them to SQL. The SS instantiates the corresponding AR 3D object to show in augmented reality on the HoloLens. In each HoloLens we have an Client App that is connected to the SS by a AR Room layer. All AR 3D objects are available to visualized by clients inside the AR Room layer.
      % These objects are synchronized with clients in the Room layer.
    \item \textbf{AR Room} - In this layer, each AR client is synchronized with the SS.
    Every AR 3D object spawned by SS can be visualized by one or more clients. Client interactions with AR objects can be visualized privately or shared with the entire network. For example, invoking an AR menu is visualized privately on the local client only, whereas spawned AR Objects resulting from the first interaction might be visualized by other connected clients.
    \item \textbf{External Device} - In this entity, external devices can connect to SS using the same or similar built applications as the SD. While SD is responsible for communicating with SS and sending SD information (such as zoom, orientation, and screen size), the new device can send queries to request data and the results in real time. This allows to users to avoid using the shared table to consult data if they wish, and the 2D application does not need to follow the same structure as the shared visualization on the SD. 
    % New devices can connect to the Shared Server and send queries to be processed and distributed among the systems. Also, the newly paired device can get processed information from multiple sources.
    \item \textbf{Query Translator API} - The query API supports and translates between different query representations (currently SQL, SPARQL, and a custom structured visual query language). This intermediate API ensures that our architecture retains the flexibility to work with different software and their particular query languages.
    
\end{itemize}

\subsection{2D Visualization}
The 2D visualization application consists of three parts, the map visualization, the visual query builder, and the text query editor where users can write their valid SPARQL queries on the query editor (see on top-right on Figure 2) and see respective query visuals on the visual query builder. Moreover, the result from both queries (visual or text) allows users to visualize selection areas on the map and their respective assets (2D objects linked to the specific latitude and longitude coordinates). The 2D visualization app supports real-time updates, which means the visual query builder, text query editor, and the map have event listeners and triggers to update the visualization during users interactions (e.g., when a user is selecting a region by visual query builder, they see the results on the visualization area instantly).
Additionally, to enable AR devices to see holographic visualization aligned with the SS, we put a QR Code in the application to support the alignment by AR Devices. AR Devices can use this QR Code as a world reference to calculate the relative size and position to align their AR content.

% The visualization application running on the SD (Figure 2) consists of three parts, the map visualization, the visual query builder, and the query editor. Users can write their valid SPARQL queries on the query editor (see on top-right on Figure 2) and see respective query visuals on the visual query builder. If the query owns the map objects (i.e., area selections), they will be visualized on the map with their respective latitude and longitude coordinates. The application supports the live updates in such a way that the visual query builder, the query editor, and the map are connected to each other in a way that any changes on any of them will be reflected on the other parts.

\begin{figure}[ht]
\label{fig:tabletopapplication}
\centering
\includegraphics[width=0.85\textwidth]{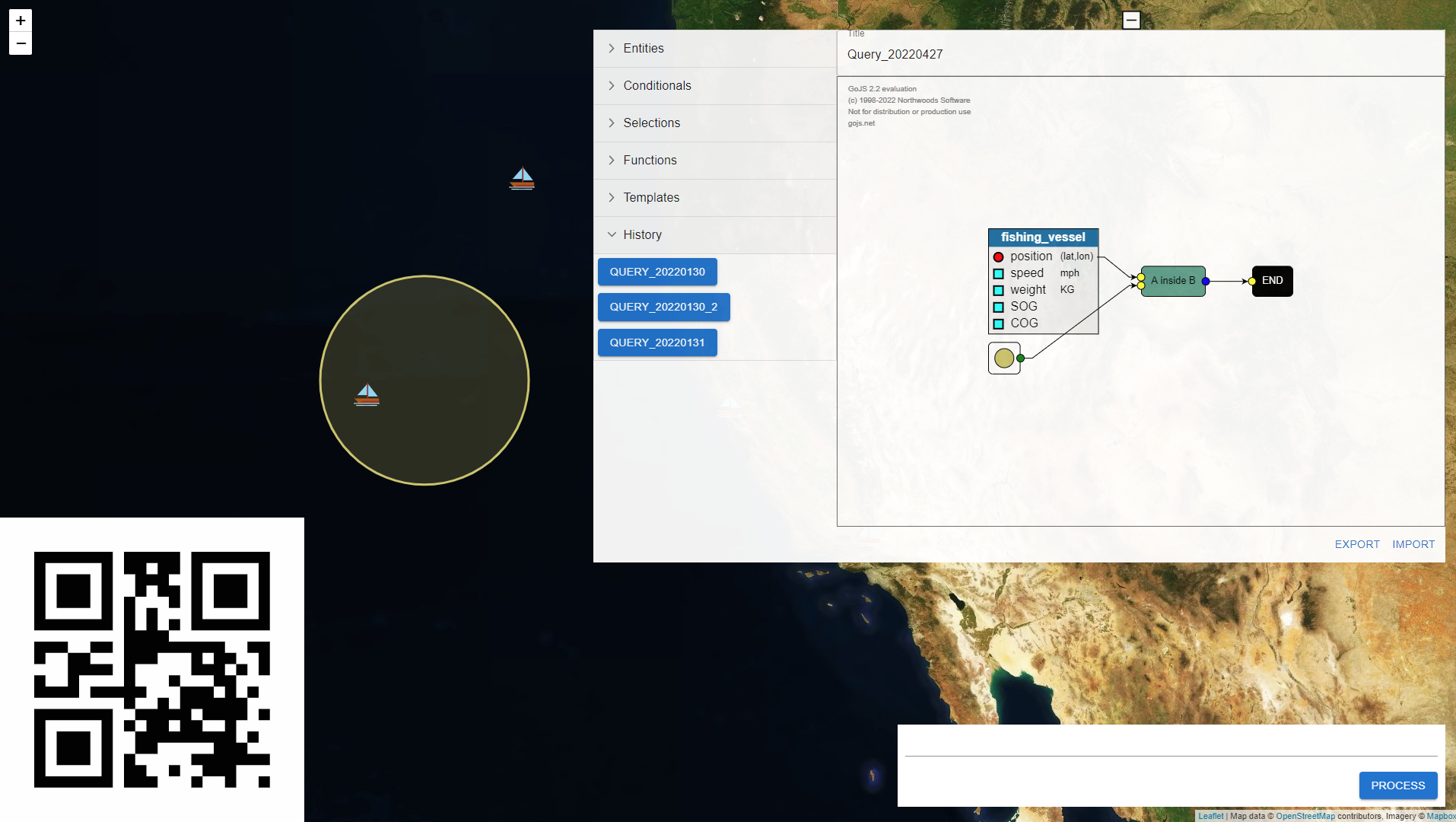}
\caption{Shared Display (SD) application developed in React. At the bottom left, a QR code is used to align the AR visualization with the SD. At the top right is the visual query builder, where the user can make data selection in real-time and see the results on the SD.}
\end{figure}

On top of the map layer (illustrated in Figure 2), we implemented an Interaction Layer (IL). On the IL, the user can interact with the map using touches, drag and drops, zoom, pans, tangible object placements, or AR interactions using AR tools (e.g., HoloLens). 

We pass direct interaction data between the SD and the 2D Application using TUIO \cite{kaltenbrunner2005tuio} over a socket connection. This protocol is commonly used for multi-touch surfaces and supports both touch and interaction using tangible objects. The IL can process information received from AR clients through their specific web-socket connection, and the updates are reflected in the visual query and the map in real-time. 

% On top of the map layer, we have considered an interaction layer. This layer is responsible for receiving all the user interactions. User interactions can be direct interactions such as map touches, drag&drops, zoom, pans, tangible object placements, or indirect interactions from augmented reality clients. The Direct interactions are in TUIO message format \cite{kaltenbrunner2005tuio} (a commonly-used protocol for the multi-touch surfaces)  that are received from the SD via a socket connection. The interaction layer also can communicate with multiple AR clients. This layer further processes the messages received from AR clients through their specific WebSocket connection, and the changes are reflected in the visual query and the map.

More detailed, to synchronize the SD application and AR clients, the 2D Application sends the visual query data, the map boundaries of the visible region in SS (north-west and south-east coordinates in latitude and longitude), and the map current zoom level to the AR clients through SS. We adopted web sockets with JSON-formatted messages  to perform this communication between SD and SS.

% In order to synchronize the SD application and AR clients, the application needs to send the visual query data, the map boundaries (north-west and south-east coordinates in latitude and longitude), and the map zoom level to AR Visualization. We have considered a WebSocket server that receives JSON messages from the table application and sends it to all the connected AR visualization to make this communication happen. 

\subsection{AR Visualization}
As outlined above, the AR application consists of two layers: a shared server (SS) and the AR Room. The SS is responsible for communicating with the database and making conversions, for example:

\begin{itemize}
    \item Converting latitude and longitude to the Unity coordinate system. We used Mapbox SDK \cite{linwood2020getting} for converting the geographic coordinate system (GCS) from our data source to Unity' coordinate system in meters.
    \item Computing coordinate system transformations between tabletop display and the AR visualization.
    \item Sending the information to Client-render layer (AR Room), which means the application responsible for generating the augmented reality in each connected AR client. 
\end{itemize} 

Second, the AR Room, which is responsible for rendering in each connected AR client (Client App) the real-time data processed to the virtual environment (including all game objects). This strategy was adopted to avoid high latency in the Client App (See on bottom-right on Figure 1). The main duties of each client App are:

\begin{itemize}
    \item Allows user to interact with objects, all interactions are computed in each client, individually.
    \item Render the content in AR.
\end{itemize}

In that sense, different AR clients can connect to the SS and they do not need to recalculate conversions that were already made by the SS. This enables the AR clients see the same virtual environment in collaborative way. 

To make this network communication between SS and AR Room, we adopt the Photon Engine SDK \cite{Realtime83:online}. Photon is a base layer for multiplayer games and higher-level network solutions. This plugin for Unity solves issues such as matchmaking and fast network communication using a scalable approach. Basically, Photon enable to create a shared-room where each client APP can connect and get the same information from server. Additionally, both (SS and Client APP) are Unity applications but with distinct duties.

In summary, each element in AR visualization are created and processed by SS, and multiple instances of Client-render connected to the SS' shared AR room can see and interact with the same AR content (Figure 3). 

\begin{figure}[ht]
\label{Fig:interacting}
\centering
\includegraphics[width=0.9\textwidth]{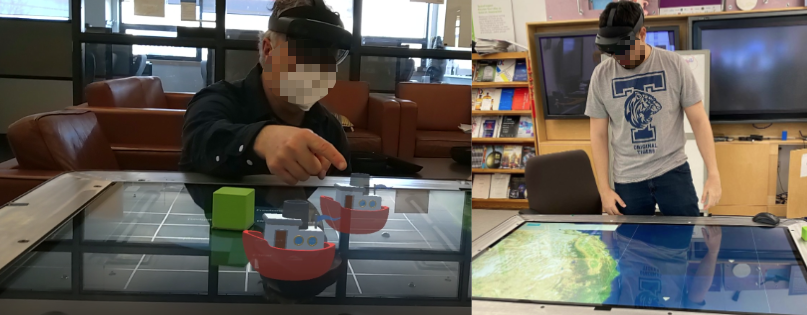}
\caption{Users are interacting with the SD wearing HoloLens. The HoloLens first-person vision of augmented reality interaction over the SD at the left. A third person of the user interacting with both systems is on the right. } 
\end{figure}

Furthermore, we implemented some interactions in an augmented reality app (Client APP) for the able user in the decision-making process during our solution. The first one is the query selection (at left in Figure 4), where the user can see an augmented reality arch connector between the world map selection on SD visualization in 2D (built in the query builder) and the 2D object that represent this selection on query builder.
In the other interaction (Figure 4, at the right), the user is grabbing an instance of an augmented reality object to show to others or get more information about the specific vessel. 

% \section{Interactions}
\begin{figure}[ht]
\label{Fig:grabbing}
\centering
\includegraphics[width=0.9\textwidth]{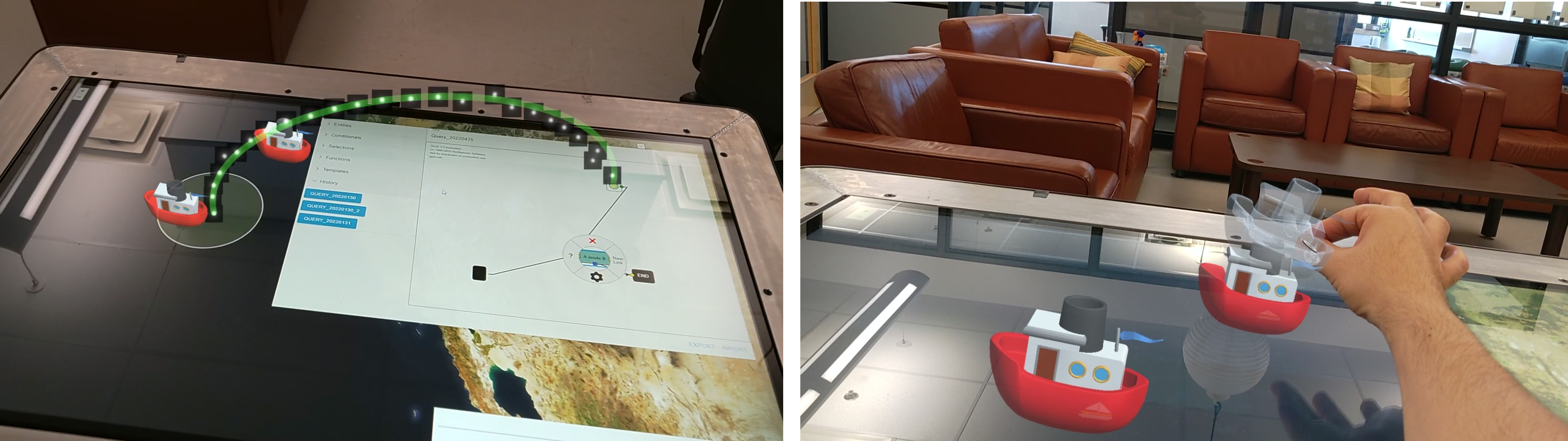}
\caption{The user selects a piece of vessel information (the user can grab and drag a the AR object to use this element to get more details about the selected vessel).}
\end{figure}

\section{Future Work and Limitations}

While the presented architecture is finished and can fit in distinct contexts, the developed prototype is under development, and there are still many challenges to overcome. Although this prototype is not a final solution, we believe this work can contribute to other researchers giving insights and directing them to develop multiple-system XR experiences for different purposes (analysis, entertainment, simulations) and fields (health, military, educational, games). 

While our study case is related to a non-entertainment context, we designed our architecture to support numerous application contexts for analysis, simulation, or entertainment applications (such as XR gaming experiences). In other words, our proposed architecture is flexible in terms of application context and field.

Besides, we are still limited in terms of developed interactions. Currently, we are working to produce more interactions among systems. While the already developed interactions can be a good study case, we realized we need to design more AR interactions before starting the usability tests phase with users. 

For this reason, the next step of this project is to conduct a user-experimental set of tests and a profound study about user behaviors in our XR interaction interfaces. We also intend to include design thinking techniques to help us construct a memorable experience for the final user of this project. Moreover, it is necessary to conduct an in-depth evaluation of users' behaviors using our solution and evaluate how this solution contributes to the decision-making process in naval organizations. 

% Another straightforward way is to explore gamification techniques to drive us to design new interactions between the systems. For this reason, the next step is to produce a multi-player augmented reality game using this proposed architecture and the related devices (such as shared displays, intelligent devices, and augmented reality head-worn displays) in an immersive gaming experience.

% In a game experience context, we can improve our application in interactive context, producing a more intuitive way. For example, this game should be explore different interactions between users, and we can analyse how is the state of enjoyment

\section{Conclusion}

We presented an architecture to integrate augmented reality with physical SD in this work. We implemented this solution to facilitate the decision-making processes in naval organizations under monitoring vessels' role. Moreover, we develop SD and AR device systems that allow multiple users to collaborate using immersive devices (e.g., HoloLens) or non-immersive such as tablets or smartphones. 

Besides, we designed and implemented different ways to visualize data (e.g., visual query builder, touchable gestures, AR gestures). In other words, our novel architecture enables multiple users to see the request and see data in an immersive or non-immersive way using particular devices (AR devices, tablets, and shared displays, which include tabletop and other touchable screens). In terms of flexibility, our architecture allows users to add or remove the immersive layer without affecting the visualization for other users. While the SD is the main non-immersive layer and is dependent on visualization, external non-immersive devices are not dependent and can be connected or disconnected at any time without affecting the visualization.

Furthermore, our proposed architecture works with a layer (Shared Server or SS) that helps to avoid unnecessary computational processing in HoloLens (concerning the limited hardware' memory and graphical processing). Moreover, the interaction of multiple users using immersive and non-immersive devices can produce rich discussion among users. 

% Additionally, we discussed that this architecture could fit in other fields, such as simulation and entertainment contexts where users interact among them. Our architecture supports further uses, which means it is appropriate for analysis, entertainment, gaming, and simulation in different fields and contexts.

We believe our architecture and prototype can help XR designers and researchers to propose new visualizations in immersive environments that combine multiple devices to facilitate decision-making processes for different purposes (simulation, education, gaming, or analysis).

% In other words, we can have users wearing augmented reality displays or without any headset (just using a SD). Additionally, our architecture is flexible in terms of visualization, where users can see data in different ways (with augmented reality or not) and collaboratively with distinct systems.
% Moreover, users can add or remove the augmented reality layer without affecting the SD visualization. Furthermore, the proposed architecture works with a layer (Shared Server or SS) that helps to avoid unnecessary computational processing in HoloLens (concerning the limited hardware' memory and graphical processing). We believe our architecture and prototype can help XR designers and researchers to propose new visualizations in immersive environments that combine multiple devices to facilitate decision-making processes for different purposes (simulation, education, gaming, or analysis). 

%Under Construction
% The AMNIS project will explore information management and utilization solutions that are relevant to the 
% future sea- and land-based military. More broadly, AMNIS will provide an avenue for academic 
% researchers in the field of machine learning, big data, graphics, and experiential media to explore science 
% issues pertinent to today’s military systems. 

\bibliographystyle{splncs04}
\bibliography{bib}

\end{document}